\documentclass[iop,apjl]{emulateapj}

\newcommand{\kms}{km\,s$^{-1}$}
\newcommand{\ms}{m\,s$^{-1}$}
\newcommand{\bz}{$\langle B_{\rm z} \rangle$}
\newcommand{\alp}{$\alpha$\,Cen\,A}
\newcommand{\eps}{$\varepsilon$\,Eri}
\newcommand{\hr}{HR\,1099}

\newcommand{\fifps}[2]{\centering\resizebox{#1}{!}{\includegraphics{#2}}}

\newcommand{\mybf}[1]{{#1}}

\shorttitle{First detection of linear polarization in cool stars}

\shortauthors{Kochukhov et al.}

\begin{document}

\title{First detection of linear polarization in the line profiles of active cool stars%
}

\author{
O.~Kochukhov\altaffilmark{1},
V.~Makaganiuk\altaffilmark{1},
N.~Piskunov\altaffilmark{1},
F.~Snik\altaffilmark{2}, 
S.V.~Jeffers\altaffilmark{2},
C.M.~Johns-Krull\altaffilmark{3},\\
C.U.~Keller\altaffilmark{2}, 
M.~Rodenhuis\altaffilmark{2}, and
J.A.~Valenti\altaffilmark{4}
}
\altaffiltext{1}{Department of Physics and Astronomy, Uppsala University, Box 516, Uppsala 75120, Sweden; oleg.kochukhov@fysast.uu.se}
\altaffiltext{2}{Sterrekundig Instituut, Universiteit Utrecht, Box 80000, 3508 TA Utrecht, The Netherlands}
\altaffiltext{3}{Department of Physics and Astronomy, Rice University, 6100 Main Street, Houston, TX 77005, USA}
\altaffiltext{4}{Space Telescope Science Institute, 3700 San Martin Dr, Baltimore MD 21211, USA}

\begin{abstract}
The application of high-resolution spectropolarimetry has led to major progress in understanding the magnetism and activity of late-type stars. During the last decade, magnetic fields have been discovered and mapped for many types of active cool stars using spectropolarimetric data. However, these observations and modeling attempts are fundamentally incomplete since they are based on the interpretation of the circular polarization alone. Taking advantage of the newly built HARPS polarimeter, we have obtained the first systematic observations of several cool active stars in all four Stokes parameters. Here we report the detection of magnetically-induced linear polarization for the primary component of the very active RS~CVn binary HR\,1099 and for the moderately active K dwarf $\varepsilon$~Eri. For both stars the amplitude of linear polarization signatures is measured to be $\sim$10$^{-4}$ of the unpolarized continuum, which is approximately a factor of ten lower than for circular polarization. This is the first detection of the linear polarization in line profiles of cool active stars. Our observations of the inactive solar-like star $\alpha$~Cen~A show neither circular nor linear polarization above the level of $\sim$10$^{-5}$, indicating the absence of a net longitudinal magnetic field stronger than 0.2~G.
\end{abstract}


\section{Introduction}
\label{intro}

Polarization in spectral lines is the most direct and reliable signature of the presence of magnetic fields on the surfaces of stars. While the Zeeman broadening of unpolarized spectra can be used, with some ambiguity, to diagnose kG-strength fields in the extremely active stars \citep{guenther:1999,johns-krull:2007}, circular polarimetry enables detection of global magnetic fields on the order of one gauss and weaker in moderately active stars \citep{auriere:2009,auriere:2010,lignieres:2009}. 

The diagnostic power of polarimetry comes from the possibility to measure very precisely magnetic signals in spectral lines against an unpolarized continuum and to boost the signal-to-noise ratio by combining weak polarization signals in thousands of spectral lines recorded simultaneously with a wide wavelength coverage \mybf{\'echelle} spectrometer \citep{donati:1997}. Furthermore, Doppler imaging analysis of the rotational modulation of circular polarization in spectral lines has resulted in detailed maps of stellar magnetic fields \citep[e.g.,][]{donati:2003}, leading to many new insights into the magnetism and activity of different groups of late-type stars \citep{donati:2009}.

\mybf{Despite} spectacular results of the recent investigations, spectropolarimetry of stellar magnetic fields remains fundamentally limited. Most importantly, high-resolution spectropolarimetric observations of all late-type stars and of the majority of early-type stars are limited to circular polarization \mybf{(Stokes $V$)}. A significantly weaker linear polarization \mybf{(Stokes $Q$ and $U$)} signal due to the transverse Zeeman effect has been detected and used for diagnosing stellar magnetic topologies only for a few strongly magnetic Ap stars \citep{wade:2000b,kochukhov:2004d} \mybf{and for one magnetic white dwarf \citep{valyavin:2008}. Previous spectropolarimetric observations of late-type stars in the Stokes $Q$ and $U$ parameters were limited to the objects where the presence of a relatively strong linear polarization was unrelated to magnetic fields \citep[e.g.,][]{boyle:1986,vink:2003}.}

Incomplete Stokes parameter data sets typically employed for magnetic stars are plagued by a reduced sensitivity to the horizontal magnetic field component and, in general, do not allow a unique reconstruction of the stellar magnetic field geometries \citep{kochukhov:2002c}. Four Stokes parameter studies of Ap stars also demonstrated that magnetic models based on the interpretation of circular polarization alone miss the smaller-scale magnetic structures \citep{kochukhov:2010}. 

\mybf{High-resolution} full four Stokes parameter observations of active cool stars represent a crucial step towards more robust and complete analysis of their magnetic field topologies. Here we present the first results of simultaneous circular \textit{and} linear polarization observations of active stars with a new polarimetric device: HARPSpol.

\section{Spectropolarimetric observations}
\label{obs}

\begin{table*}
\centering
\small
\caption{
Journal of the HARPSpol four Stokes parameter observations of late-type stars. 
}
\label{tbl:obs}
\begin{tabular}{ccccccccrc}
\hline
\hline
Name & Date  & HJD             & Stokes & $t_{\rm exp}$& S/N & $\sigma_{\rm LSD}$ & $P_{\rm max}$ & \multicolumn{1}{c}{FAP} & Det. \\
           &  (UT)  & (2400000+) & parameter & (s)            &        & ($10^{-5}I_{\rm c}$) & ($10^{-5}I_{\rm c}$)  & & flag \\
\hline
\alp & 02 Jun 2009 & 54984.6493 & $V$ & $72\times10$ & 3410 & 0.87 & $<$1.92 & 9.9E-01 & ND\\
     & 02 Jun 2009 & 54984.6915 & $Q$ & $72\times10$ & 2630 & 0.69 & $<$1.34 &  6.9E-01 & ND\\
     & 02 Jun 2009 & 54984.7380 & $U$ & $96\times10$ & 3010 & 0.60 & $<$1.19 & 9.8E-01 & ND\\
\hline
\hr  & 04 Jan 2010 & 55200.6197 & $V$ & $4\times421$ & 390 & 4.88 & 190.9& $<$1.0E-16 & DD\\
     & 04 Jan 2010 & 55200.6441 & $Q$ & $4\times421$ & 360 & 2.88 & 15.1 & $<$1.0E-16 & DD\\
     & 04 Jan 2010 & 55200.6715 & $U$ & $4\times421$ & 370 & 2.81 & 11.0 & 7.2E-08 & DD\\
     & 16 Jan 2010 & 55212.5897 & $V$ & $4\times421$ & 440 & 4.24 & 92.0 & $<$1.0E-16 & DD\\
     & 16 Jan 2010 & 55212.5458 & $Q$ & $4\times421$ & 420 & 2.44 & 12.1 & $<$1.0E-16 & DD\\
     & 16 Jan 2010 & 55212.5675 & $U$ & $4\times421$ & 420 & 2.44 & 11.6 & 1.8E-14 & DD\\
\hline
\eps & 08 Jan 2010 & 55204.5758 & $V$ & $4\times250$ & 880 & 1.99 & 16.4 &  $<$1.0E-16 & DD\\
     & 08 Jan 2010 & 55204.5434 & $Q$ & $4\times250$ & 880 & 1.20 & 4.47 &  2.7E-05 & MD\\
     & 08 Jan 2010 & 55204.5605 & $U$ & $4\times250$ & 880 & 1.15 & 2.42 &  8.2E-01 & ND\\
     & 09 Jan 2010 & 55205.6000 & $V$ & $4\times240$ & 820 & 2.22 & 46.9 &  $<$1.0E-16 & DD\\
     & 09 Jan 2010 & 55205.5714 & $Q$ & $4\times240$ & 860 & 1.22 & 3.85 &  1.9E-05 & MD\\
     & 09 Jan 2010 & 55205.5853 & $U$ & $4\times240$ & 820 & 1.25 & 5.49 &  2.0E-05 & MD\\
     & 11 Jan 2010 & 55207.6232 & $V$ & $4\times240$ & 770 & 2.39 & 33.0 &  $<$1.0E-16 & DD\\
     & 11 Jan 2010 & 55207.5949 & $Q$ & $4\times240$ & 800 & 1.32 & 5.81 &  1.4E-10 & DD\\
     & 11 Jan 2010 & 55207.6086 & $U$ & $4\times240$ & 780 & 1.33 & 2.81 &  7.9E-01 & ND\\
\hline
\end{tabular}
\tablecomments{ND, MD, and DD in the last column indicate no, marginal, and definite detections of polarization signature.}
\end{table*}

The HARPS spectrometer \citep{mayor:2003}, fed by the 3.6-m telescope at the European Southern Observatory (ESO), is one of the most powerful astronomical tools for precision studies of stellar spectra.
It is the highest resolution ESO spectrograph, providing a coverage of the 380--690~nm wavelength region in a single exposure with the resolving power of $\lambda/\Delta\lambda$\,=\,115,000 and a long-term velocity stability of better than 1~\ms. 

The new polarimeter, HARPSpol \citep{snik:2010,piskunov:2011}, is installed in the Cassegrain focus of the 3.6-m telescope, feeding the two existing HARPS optical fibers. The polarimeter consists of two separate polarimetric units: one for circular and another for linear polarization measurements. Each unit is equipped with a modulator -- a superachromatic quarter-wave or half-wave plate. The light is split into two orthogonally polarized beams by a polarizing beam-splitter (Foster prism). The corresponding spectra are recorded simultaneously on the \mybf{4\,k$\times$4\,k} CCD mosaic. A slider inserts the polarimeter from a completely retracted position, and allows for selection of either the circular or the linear polarimeter.

The two orthogonally polarized beams injected into the fibers can be exchanged by rotating the half-wave plate with the increment steps of 45$^{\rm o}$ and the quarter-wave plate with 90$^{\rm o}$ steps. A combination of exposures obtained with different modulator orientations allows compensation of, to first order, flat-fielding errors and effects of seeing and other artifacts \citep{semel:1993}.
The high spectral resolution of \mybf{HARPSpol} is especially useful for detecting complex and weak linear polarization signatures in cool stars \citep{semel:2009}.

Each of our polarization observations of cool stars consisted of four sub-exposures, which were calibrated and processed with the help of the REDUCE package \citep{piskunov:2002}. The resulting one-dimensional spectra were demodulated using the ``ratio'' method described by \citet{bagnulo:2009} and re-normalized to the continuum intensity, yielding a spectrum in one of the Stokes parameters ($V/I_{\rm c}$, $Q/I_{\rm c}$ or $U/I_{\rm c}$), the corresponding intensity spectrum ($I/I_{\rm c}$), associated error bars, and a diagnostic null spectrum. The circular to linear polarization cross-talk is estimated to be below 0.5\% using symmetry properties of the  Stokes profiles for the slowly rotating Ap star $\gamma$~Equ.

Table~\ref{tbl:obs} summarizes the four Stokes parameter observations of three late-type stars -- \hr, \eps, \alp\ -- obtained during our HARPSpol runs in May 2009 and January 2010. The first six columns in that table give the target name, the observation date, the heliocentric Julian date, \mybf{the $V$, $Q$, and $U$ Stokes parameters}, the exposure time, and the peak signal-to-noise ($S/N$) ratio per 0.8~\kms\ velocity bin.

\begin{figure*}[!th]
\fifps{5.6cm}{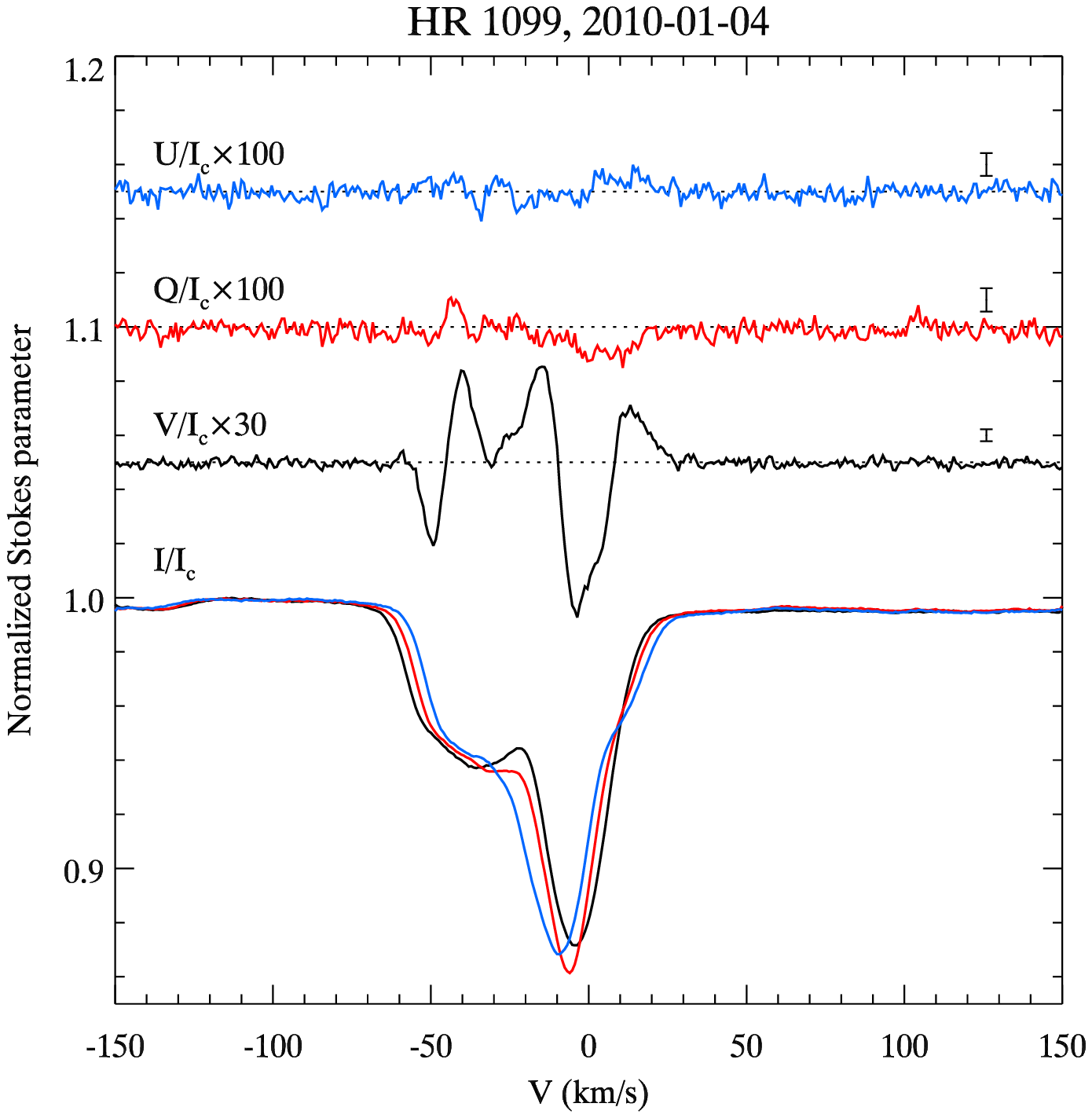}		
\fifps{5.6cm}{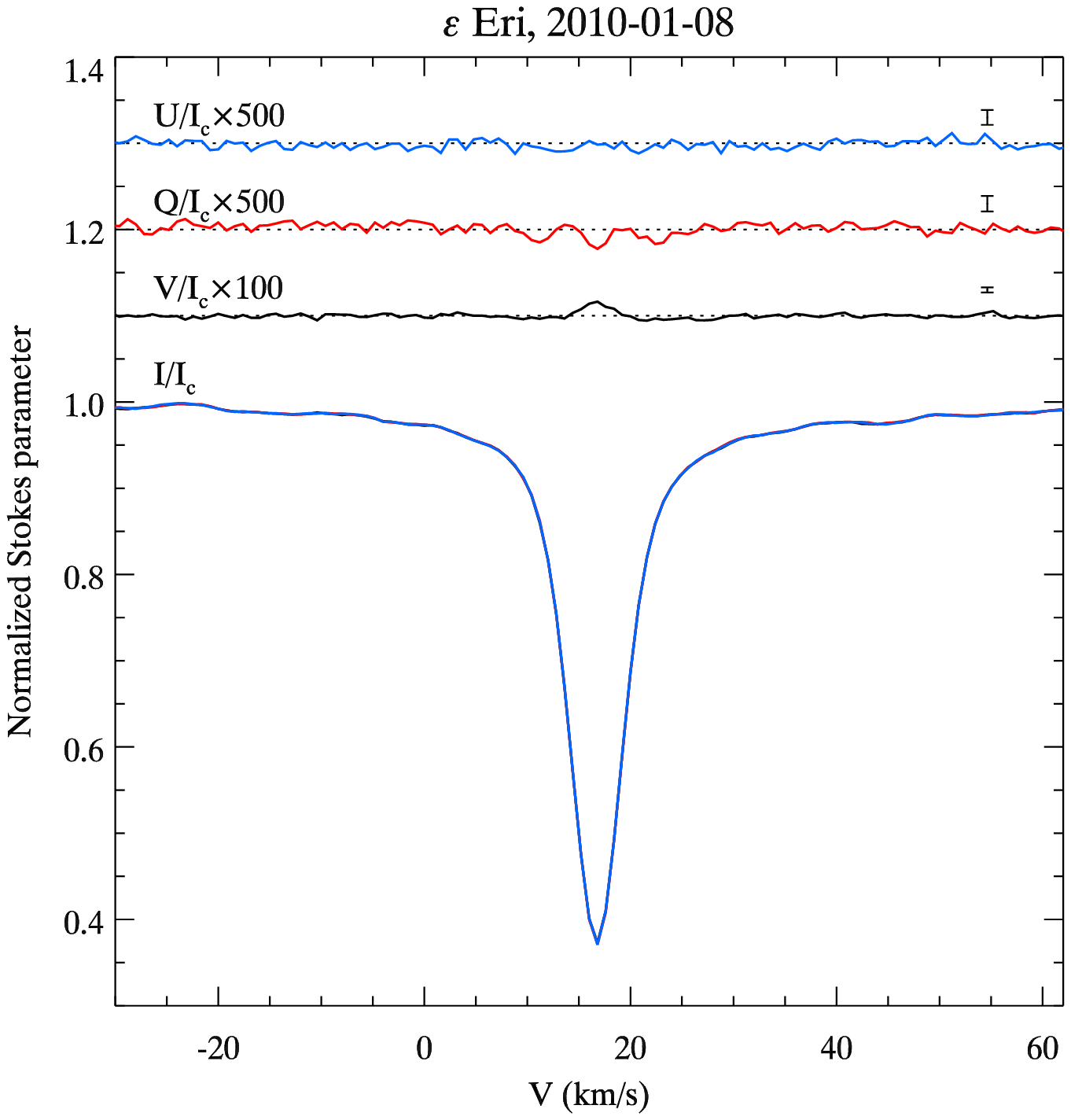}		
\fifps{5.6cm}{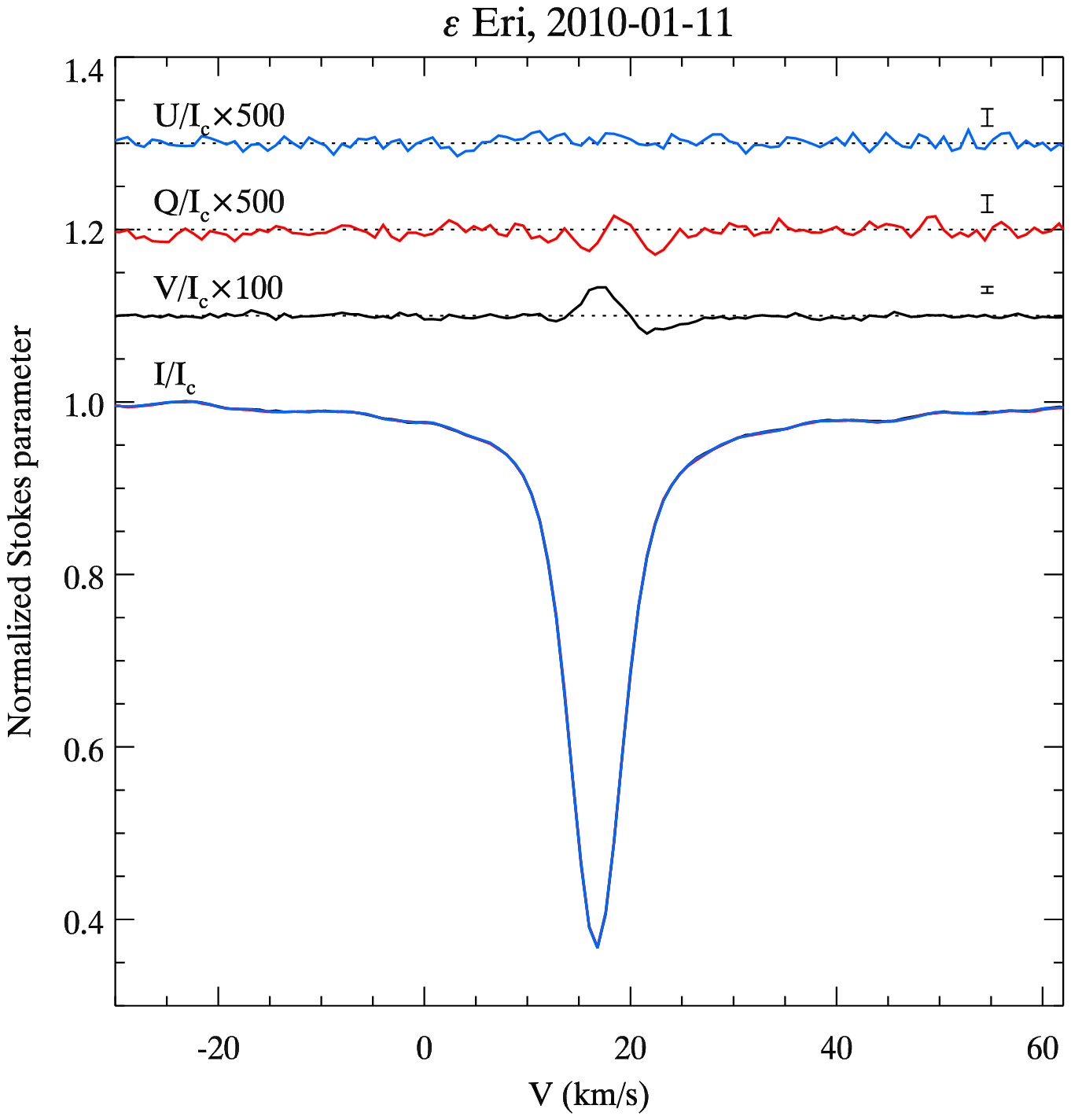}		
\fifps{5.6cm}{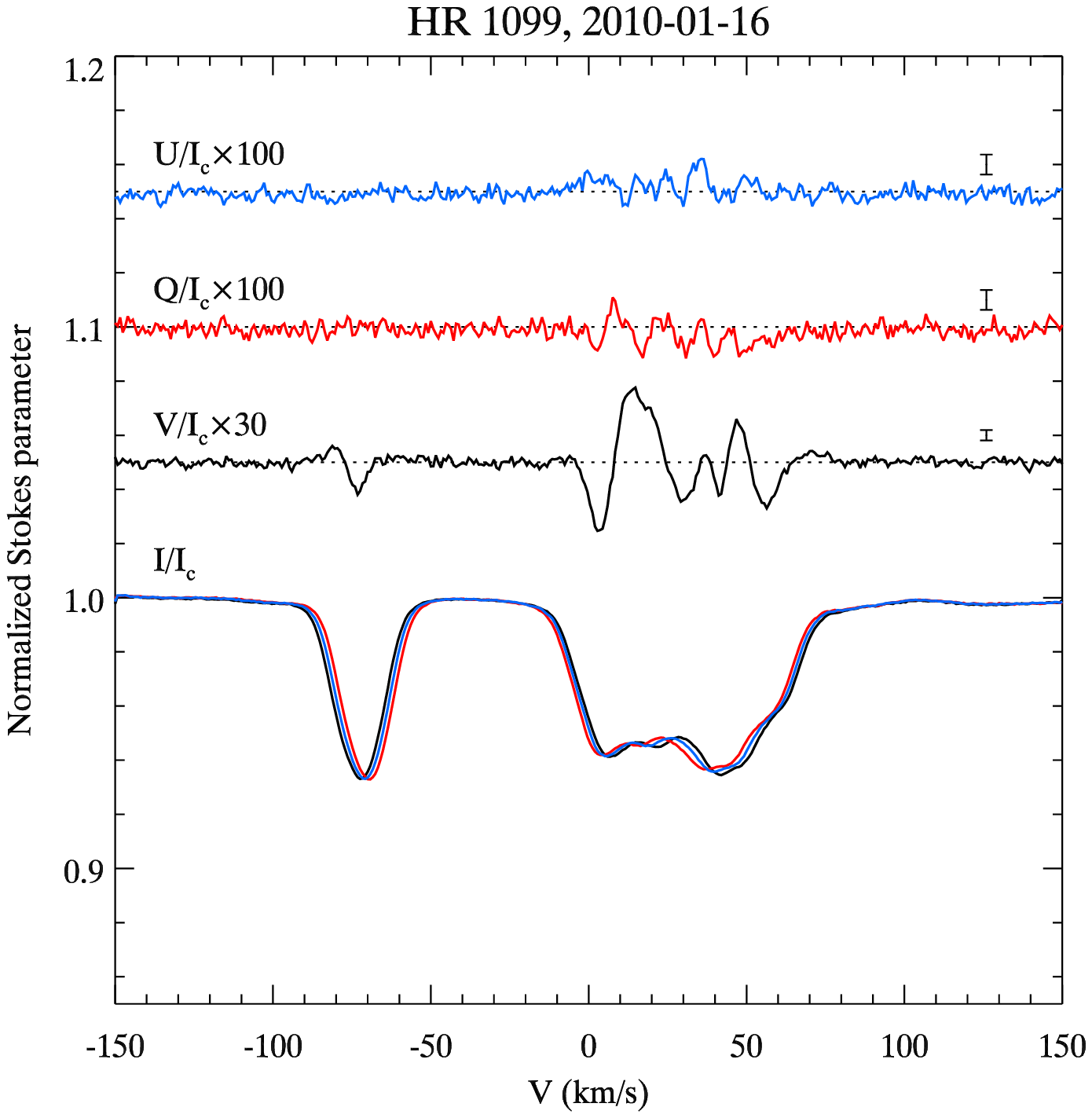}		
\fifps{5.6cm}{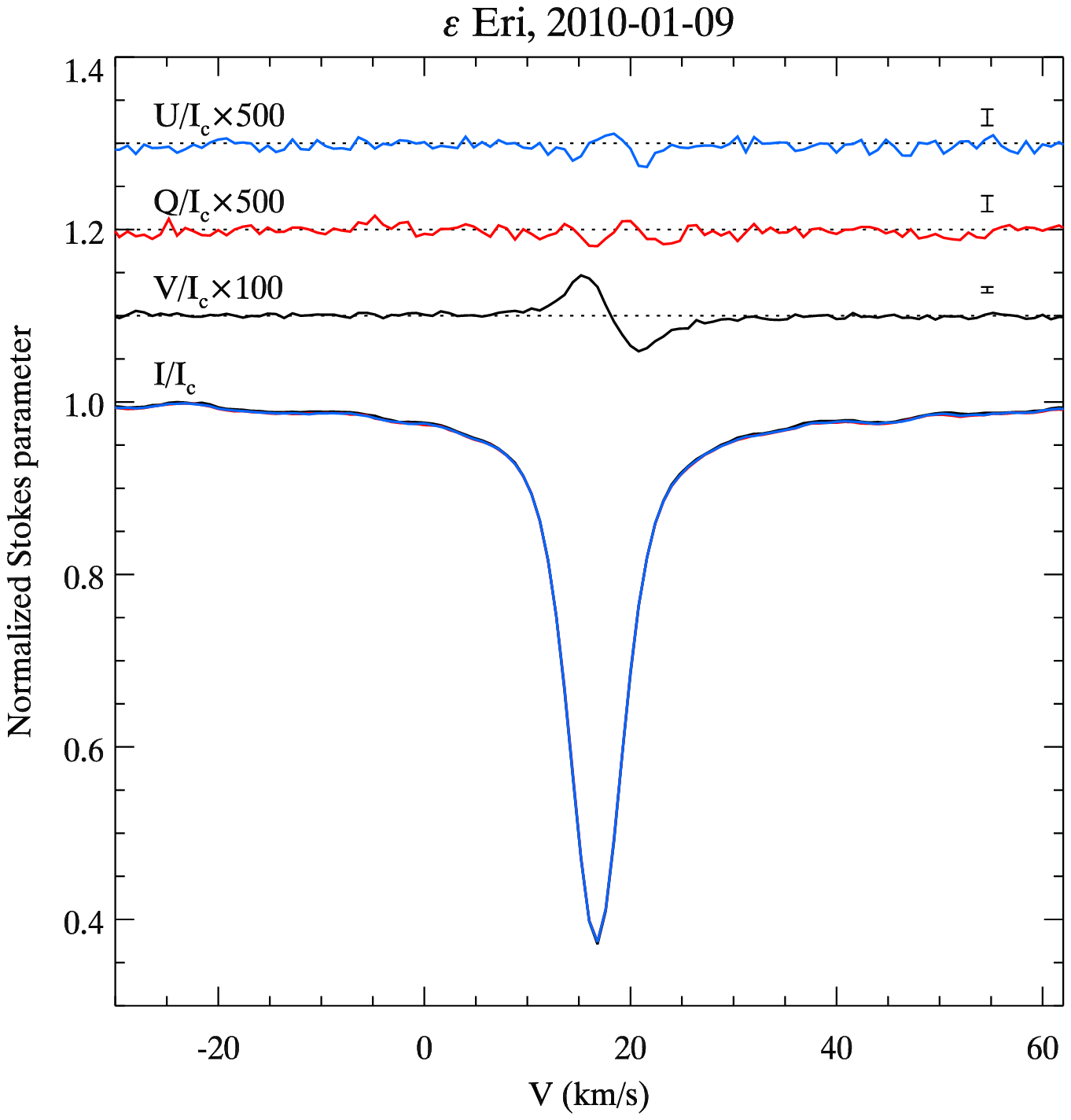}		
\fifps{5.6cm}{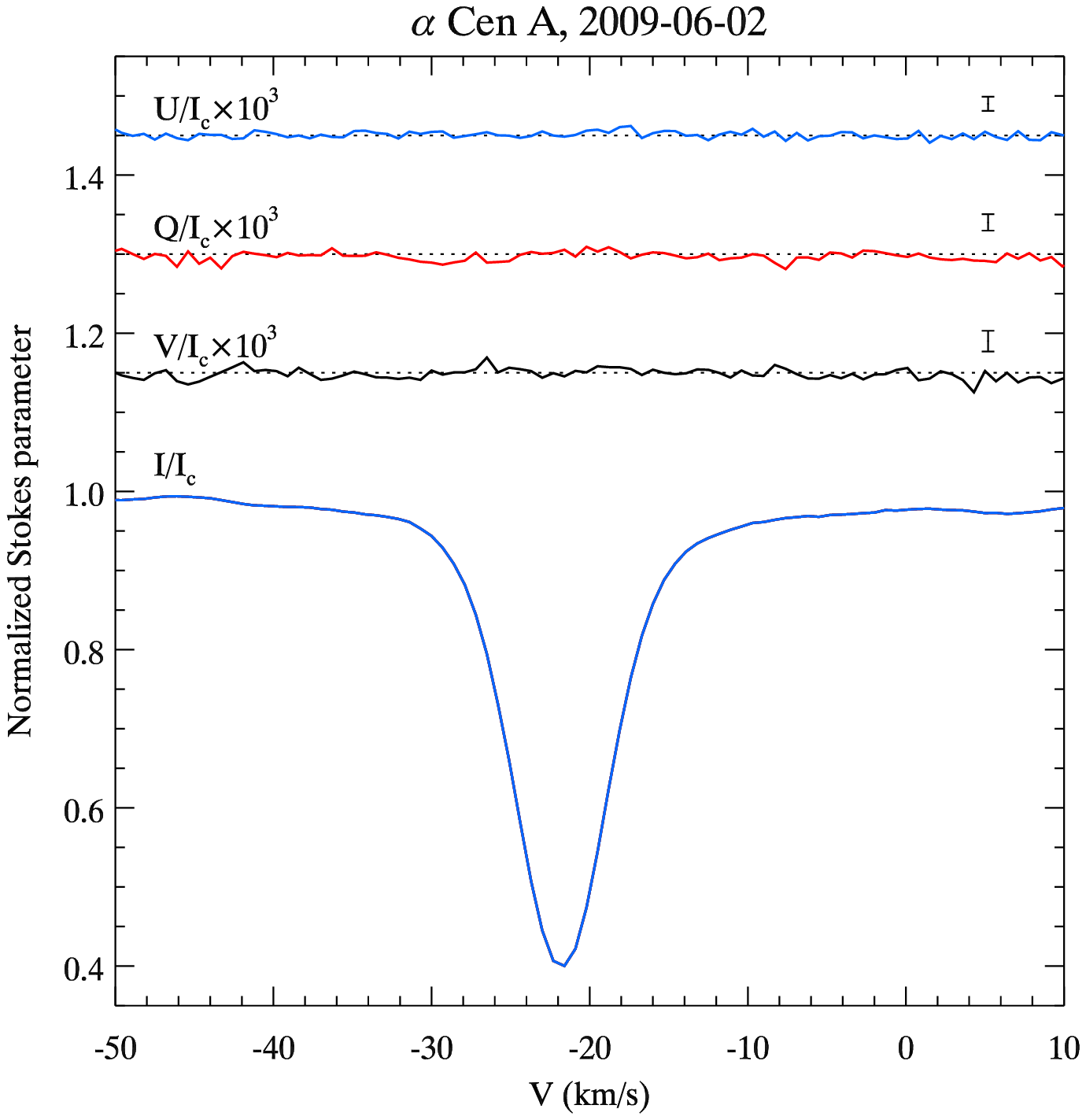}		
\caption{
Four Stokes parameter LSD profiles derived from the HARPSpol observation of \hr, \eps, and \alp. Polarization profiles are shifted vertically and expanded relative to the mean intensity profile. The bars on the right correspond to \mybf{three} times the noise level of the LSD polarization profiles.
}
\label{fig:lsd}
\end{figure*}

\section{Multi-line polarization analysis}
\label{lsd}

Polarization signatures are typically not detectable in individual spectral lines of active late-type stars. Among the targets observed in our program, only \hr\ shows evidence of the circular polarization signal in a few strong, magnetically sensitive spectral lines. To detect and characterize magnetic fields of the target stars we use a multi-line technique known as least-squares deconvolution \citep[LSD,][]{donati:1997}. This method approximates the stellar spectrum with a superposition of identical profiles, centered at the position of individual lines and scaled according to the line's strength and magnetic sensitivity. A mean profile extracted by LSD combines common information from thousands of spectral lines, boosting the sensitivity of polarization studies.

We computed mean profiles for all Stokes parameters and the corresponding null polarization spectra using the methodology and the code developed by \citet{kochukhov:2010a}. The LSD line mask was based on the atomic line parameters extracted from the \mybf{Vienna Atomic Line Database} \citep[VALD,][]{kupka:1999}. Depending on the star, we used between 7500 and 9000 metal lines stronger than 10\% of the unpolarized continuum, achieving a multiplex gain of 30--100.

Table~\ref{tbl:obs} summarizes results of the LSD analysis of our targets. Columns 7--10 give information on the noise level (per 0.8~\kms\ velocity bin) in the LSD profiles, the maximum amplitude of the LSD polarization signal, the false alarm probability (FAP) of the signal detection computed according to \citet{donati:1992}, and the polarization detection flag. Following the widely used spectropolarimetric convention, we consider profiles with FAP\,$<10^{-5}$ as a definite detection (DD), $10^{-3}<$\,FAP\,$<10^{-5}$ as a marginal detection (MD), and FAP\,$>10^{-3}$ as no detection (ND).

\section{Results}
\label{results}

\subsection{\hr}

The K1 subgiant primary of the RS CVn binary \hr\ (V711~Tau) is one of the most magnetically active late-type stars. It was frequently studied with Doppler imaging \citep{vogt:1999} and circular spectropolarimetry \citep{donati:2003}. These investigations found cool spots evolving on a time scale of about one year and longer-lived magnetic field structures, dominated by regions with predominantly azimuthal field orientation.

We observed HR\,1099 in all four Stokes parameters on two different nights in January 2010. Fig.~\ref{fig:lsd} shows the LSD Stokes $IQUV$ profiles obtained for this star. \mybf{The contribution of both stars is evident in Stokes $I$.} The distortions in the mean intensity profile of the primary indicate significant spot coverage at the time of our observations. The rapid changes of Stokes $I$ are due to the orbital motion and rotation of the primary with $P$\,$\approx$\,$2.838$~d \citep{fekel:1983}. The Stokes $V$ profile of the primary shows a complex field morphology. Using the first moment of LSD Stokes $V$ \citep{kochukhov:2010a}, we derive a mean longitudinal magnetic field of \bz\,=\,$25.2\pm3.6$~G and $12.3\pm3.0$~G for the observations on January 4 and 16, respectively. A weak magnetic signature in the Stokes $V$ profile of the secondary component is also clearly detected \mybf{at $V\approx-70$~\kms\ } in the latter observation.

The $Q$ and $U$ signatures of the primary are detected in both HARPSpol observations with a very high confidence (FAP\,$<10^{-7}$). This is the first detection of the linear polarization in line profiles of a cool active star. The LSD Stokes $QU$ profiles of \hr\ are significantly more complex than the corresponding Stokes $V$ spectra and have an amplitude of $\sim$$10^{-4}I_{\rm c}$. No signal is present in the $Q$ and $U$ null profiles (FAP\,$>0.8$), confirming that detected linear polarization is intrinsic to the star and does not arise due to an instrumental or data reduction artifact.

\subsection{\eps}

The K2 dwarf \eps\ is one of the nearest solar-type stars and the closest exoplanet host \citep{hatzes:2000,benedict:2006}. This star is surrounded by a debris disk and shows a high level of magnetic activity. Previous studies of the Zeeman broadening in optical and infrared magnetically sensitive lines suggested the presence of fields with spatially averaged strength of 100--200~G \citep{valenti:1995,rueedi:1997}. No detection of the magnetically induced line polarization was previously reported for this star.

We observed \eps\ in all four Stokes parameters during 11 nights, fully covering the 11.3~d rotation period of the star \citep{frohlich:2007}. Fig.~\ref{fig:lsd} illustrates the LSD intensity and polarization spectra of \eps\ for the three selected nights. There is no evidence of variability in Stokes $I$, but a variable magnetic field is detected in the circular polarization for all nights. 
The Stokes $V$ LSD profiles change systematically from night to night throughout the rotation period of the star.
The mean longitudinal magnetic field of \eps\ varies between $-5.8\pm0.1$ and $4.7\pm0.1$~G.

Despite a high $S/N$ ratio of our polarimetric data, the LSD $Q$ and $U$ profiles of \eps\ generally do not reveal magnetic signatures. However, on the night 11 January 2010 the $Q$ signature is definitely detected, while on the other two nights a marginal signal may be present in $Q$ and/or $U$ profiles. 
The amplitude of these linear polarization signatures does not exceed $6\times10^{-5} I_{\rm c}$. \mybf{Rotational modulation of the linear polarization signal and its occurrence within the stellar photospheric lines suggests that polarization is unrelated to the dusty disk around \eps.}

\subsection{\alp}

The solar analogue \alp\ was observed with \mbox{HARPSpol} in June 2009, during partly cloudy conditions. Nevertheless, sufficient photons were collected to reach a polarimetric sensitivity of $\sim$\,$10^{-5}$ with the help of the LSD technique. As demonstrated by Fig.~\ref{fig:lsd}, no signals were detected in either $V$, $Q$ or $U$ profiles. The corresponding null spectra are also featureless, confirming the lack of polarization artifacts down to the level of $\sim$\,$10^{-5}$. 

The non-detection of the Stokes $V$ signature corresponds to the $3\sigma$ upper limit of 0.2~G for the disk-averaged longitudinal magnetic field. In comparison, the net longitudinal field of the Sun as a star reaches 1~G during activity maximum and is below 0.1~G during the minimum \citep{demidov:2002}. We therefore conclude that HARPSpol can detect the global magnetic fields of active solar twins. In its present low activity state, \alp\ appears to be reminiscent of the Sun at cycle minimum, in agreement with recent X-ray observations \citep{ayres:2009}.

\section{Discussion}

For both \hr\ and \eps\ we detected linear polarization in spectral lines with an amplitude of 0.6--15$\times10^{-4}$ of the unpolarized continuum, which is about a factor of 10 lower than the amplitude of the circular polarization signal recorded on the same nights. For both objects the LSD linear polarization profiles appear to be morphologically complex, requiring resolving power $\lambda/\Delta\lambda\sim10^5$ for secure detection.

It is interesting to discuss the cause of linear polarization in cool stars. 
\mybf{Solar-like spots are connected with a patchy, magnetic environment called ``plage'' through low-lying magnetic loops \citep{solanki:2006}. These plage regions are usually brighter than the quiet photosphere and much brighter than the cool spots, and therefore dominate the disk-integrated magnetic signal. For an on-disk active region, linear polarization can be created if the plage is not symmetrically distributed around the spot. Near the limb, the vertical components of plage can create significant linear polarization as long as there are not too many active regions such that their magnetic signals average out over the disk.} The two polar regions that exhibit a weak but quite uniform outward magnetic field over a large area \citep{tsuneta:2008} could contribute to linear polarization as well.

On the other hand, for stars with enhanced activity and non-solar dynamo properties, such as RS CVn binaries, linear polarization can arise from the large-scale toroidal fields mapped by circular polarization \mybf{Zeeman Doppler imaging} studies \citep{donati:2003}. Four Stokes parameters observations with complete rotational phase coverage are essential to confirm the reality of these azimuthal fields and distinguish different scenarios of their formation.

\acknowledgements
OK is a Royal Swedish Academy of Sciences Research Fellow supported by grants from the Knut and Alice Wallenberg Foundation and the Swedish Research Council.


\end{document}